\documentclass[twocolumn,showkeywords,aps,pra,superscriptaddress,tightenlines,article]{revtex4}
\usepackage{amsmath}
\usepackage{amsfonts}
\usepackage{graphicx}
\usepackage{epsfig}

\usepackage{color}
\usepackage[dvipsnames,svgnames,x11names]{xcolor}
\usepackage[colorlinks,citecolor=blue]{hyperref}

\let\oldtextcolor\textcolor
\renewcommand{\textcolor}[2]{\oldtextcolor{black}{#2}}
\begin{document}

\title{Preparing Spin Squeezed States via Adaptive Genetic Algorithm}

\author{Y. M. Zhao}
 \affiliation{School of Physics, Qingdao University of Technology, 0532, Qingdao, Shandong, China}
\author{L. B. Chen}
 \affiliation{School of Physics, Qingdao University of Technology, 0532, Qingdao, Shandong, China}
 \author{W. Z. Zhang}
 \affiliation{Department of Physics, School of Physical Science and Technology, Ningbo University, 315211, Ningbo, Zhejiang, China}
\author{Y. Wang}
 \affiliation{School of Physics, Qingdao University of Technology, 0532, Qingdao, Shandong, China}
\author{H. Y. Ma\footnote{Corresponding author\quad E-mail:~\textsf{mahongyang@qut.edu.cn}}}
 \affiliation{School of Physics, Qingdao University of Technology, 0532, Qingdao, Shandong, China}
\author{X. L. Zhao\footnote{Corresponding author\quad E-mail:~\textsf{zhaoxiaolong@qut.edu.cn}}}
 \affiliation{School of Physics, Qingdao University of Technology, 0532, Qingdao, Shandong, China}
\date{\today}

\begin{abstract}
We introduce a novel strategy employing an adaptive genetic algorithm (GA) for iterative optimization of control sequences to generate
quantum nonclassical states. Its efficacy is demonstrated by preparing spin-squeezed states in an open collective spin model governed
by a linear control field. Inspired by Darwinian evolution, the algorithm iteratively refines control sequences using crossover, mutation,
and elimination strategies, starting from a coherent spin state within a dissipative and dephasing environment. 
\textcolor{red}{We rigorously benchmark our method against constant control protocols and reinforcement learning, demonstrating competitive and robust performance. Furthermore, we showcase the GA's versatility by directly optimizing for metrologically relevant squeezing, achieving scalable performance, even in the presence of dissipation and thermal noise. The proposed strategy demonstrates a high state-preparation fidelity, exceeding 0.99, and provides a long time window for maintaining the spin squeezed state, even under dissipative conditions.}
 We discuss feasible experimental implementations and potential
extensions to alternative quantum systems, and the adaptability of the GA module. This research establishes the foundation for utilizing
GA-like strategies in controlling quantum systems and achieving desired nonclassical states.
\end{abstract}

\maketitle
\section{Introduction}
Quantum-enhanced metrology has emerged as a cornerstone in the quest for ever-greater precision in measurement, pushing the boundaries of what
is achievable using classical systems. Central to this advancement is the development and manipulation of nonclassical quantum states, such as
spin-squeezed states, which can surpass the standard quantum limit~\cite{1PR50989,2RMP90035005}.
Spin squeezing often arises concurrently with entanglement, a consequence of nonlinear interactions within the ensemble~\cite{1PR50989,2RMP90035005}.
When properly harnessed, this phenomenon has the potential to revolutionize precision measurements by lowering quantum noise in targeted directions
of the system's state space.
These states reduce quantum uncertainty in specific components of a system's collective spin, enabling unprecedented sensitivity in measurements
that are crucial for applications ranging from homodyne interferometers~\cite{3PRA46R6797,4PRA5067,5nature529505} to optical atomic clocks~\cite{8nature621734,9nature20208,10nature588414} and magnetometers~\cite{6PRL109253605}. However, generating and maintaining highly squeezed
states in practical, noisy environments remains a formidable challenge, particularly in systems
subject to dissipation.

Various approaches have been proposed to achieve spin squeezing~\cite{1PR50989,2RMP90035005,12PRA475138}. These include quantum non-demolition
measurements~\cite{13EL42481}, coherent control~\cite{14PRA63055601}, and the use of nonlinear interactions in systems such as Bose-Einstein condensates (BECs)~\cite{16Nature40963,17Nature4641165,18Nature4641170}. While significant progress has been made, there is a continuing need for control methods
that not only generate strong squeezing but also maintain coherence over extended periods, especially in the presence of environmental noise.

Machine learning techniques, particularly reinforcement learning (RL) \textcolor{red}{and variational quantum circuits (VQCs)}, have recently been applied to optimize control strategies in high-dimensional
quantum systems~\cite{19Murphy2012,20Sutton2018,21Nature549195,nrp3625,prl123260505,prap17064050,prr5043285,prr7l022072}. These methods excel at exploring complex control landscapes, but they often come
with high computational costs and complex hyperparameter tuning, making them less feasible for real-time or large-scale applications. This has spurred
interest in alternative optimization strategies that are both
scalable and easier to implement. Evolutionary strategies, particularly GAs, offer a promising alternative~\cite{22arx170303864}.
GAs simulate the process of natural selection, iteratively refining a population of candidate solutions through mechanisms such as crossover, mutation,
and selection. In the context of quantum control, GAs can be particularly advantageous: they offer a flexible and adaptable framework for optimizing
control sequences, requiring less detailed prior knowledge of the system's dynamics than gradient-based approaches. Additionally, GAs are well-suited
for discrete control problems, such as the optimization of sequences of control pulses~\cite{CEP1011}, where they can explore a vast parameter space with
high efficiency. \textcolor{red}{Furthermore, as we will demonstrate, they can achieve performance competitive with more complex machine learning paradigms.}

Building upon these foundations, we propose a novel GA-based optimal control strategy for preparing non-classical spin states in an open environment.
Our approach leverages a sequence of square pulses, mimicking bang-bang control~\cite{26PRA582733,27PRA86022321}, to steer the system towards a desired
squeezed state. Optimization is achieved through a process mirroring natural selection, where candidate control sequences, encoded as `individuals', undergo
crossover and mutation within `populations'. Through iterative generations, the fittest individuals are selected, driving the system towards the optimal
control solution. We rigorously evaluate the performance of this scheme across a wide range of control parameters, system sizes, and thermal environments, demonstrating its robustness and efficacy. \textcolor{red}{We go beyond a simple demonstration by thoroughly benchmarking our algorithm against both a constant control scheme and a state-of-the-art RL agent, highlighting the GA's competitive performance. We further showcase the versatility of our framework by directly optimizing for metrologically relevant squeezing, revealing that the GA can discover control protocols that yield scalable squeezing approaching the fundamental Heisenberg limit, even under dissipative conditions.} Furthermore, we explore the scalability of the algorithm and its potential applicability to other quantum systems,
such as BECs, highlighting its experimental feasibility.

The structure of this paper is as follows: In Section \ref{ConStr}, we introduce the GA-based optimization framework designed for the preparation of
nonclassical quantum states. Section \ref{RLandGA} details the integration of the GA module within the control scheme. In Section \ref{CSM}, we describe
the quantum model employed for generating target spin-squeezed states. Section \ref{PSSS} outlines the procedure for preparing spin-squeezed states in
an open collective spin system using the adaptive GA. In Section \ref{ConRes}, we evaluate the performance of the proposed method, analyzing the impact
of control pulse frequency, control type diversity, system scalability, and robustness. Section \ref{discussion} discusses potential experimental
implementations, \textcolor{red}{and compares GA with alternative control strategies}. The adaptability and scalability of the GA
module are also discussed. Section \ref{CONC} summarizes the findings of this study.

\section{Control Scheme}\label{ConStr}

Inspired by Darwin's theory of evolution~\cite{28darwin}, genetic algorithms (GAs) have emerged as powerful optimization techniques based on the principles
of natural selection. GAs simulate evolutionary processes, subjecting a population of candidate solutions to selection, crossover, and mutation to
iteratively generate improved offspring~\cite{29holland1992}, which effectively navigates complex search spaces. This powerful framework has found widespread
application across diverse fields, from engineering design to financial modeling, demonstrating itsefficacy in tackling complex, non-linear
problems~\cite{30goldberg1987}.

Leveraging the generality of GAs, we propose a control scheme utilizing an adaptive GA to optimize the arrangement of control pulses for the preparation
of nonclassical states. We consider a quantum system governed by a general Hamiltonian under the influence of control fields:
$\hat{H}=\hat{H}_0+\sum_{k=1}^{K}f_k(t)\hat{H}_k$, where $\hat{H}_0$ is the free Hamiltonian and $K$ denotes the amount of the external control Hamiltonians
$\hat{H}_k$. $f_k(t)$ is the amplitude of
a time-dependent control field, designed by the GA algorithm. Because of the Heisenberg uncertainty principle, it should be confirmed
that $[\hat{H}_{0},\hat{H}_{k}]\neq0$, otherwise, the influence of the control Hamiltonians can be subsumed into the free Hamiltonian.

\begin{figure}
 \includegraphics*[width=8cm]{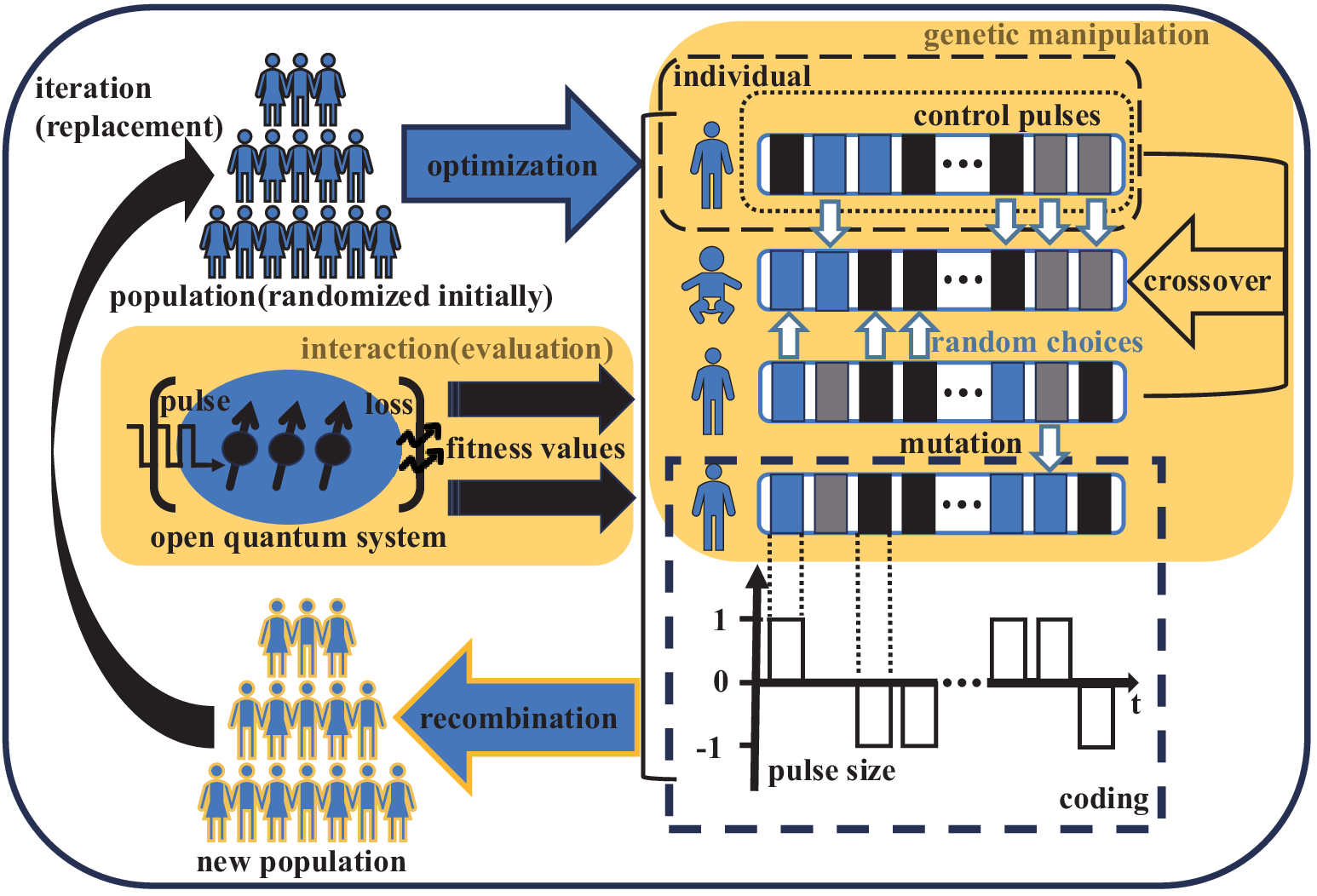}
 \caption{The GA scheme is used to optimize control pulses for squeezing a spin system in an open quantum system. It starts with a randomly generated
 population of individuals, each representing a unique control pulse sequence. Individuals undergo crossover and mutation operations to create new offspring.
 Fitness is evaluated by simulating the open quantum system dynamics under each control pulse, with higher squeezing leading to greater fitness. The GA
 iteratively selects individuals based on their fitness, promoting the propagation of optimal pulse sequences across generations until the set number of
 iteration limit is reached.}
 \label{RLPro}
\end{figure}

FIG~\ref{RLPro} illustrates the process of obtaining the control sequence by closed-loop emulation, whereas it is implemented in an open-loop control process.
The control sequence will steer the system towards a set of states satisfying the desired control target. Our proposed GA framework encodes each individual
control sequence within a population that may include the optimal solution. The space of the control sequence explodes exponentially versus the number of
control pulses. This results in a huge space for the population.

The versatility of this GA-based approach extends beyond optimizing pulse sequences. It can be readily adapted to optimize the other parameters of the system
crucial for achieving the target set. Furthermore, this framework possesses inherent generalizability to be applied to other dynamical systems governed by
differential equations. To showcase its efficacy, we demonstrate its application in engineering a collective spin system, illustrating its potential for precise
control and manipulation of quantum states.

\section{Genetic Algorithm optimizes Control Sequence}
\label{RLandGA}

This section details the logic underpinning our adaptive GA-based optimization strategy for maximizing spin squeezing which can be broadly divided into
three key parts:

1. Encoding: GA commences by generating an initial population \begin{equation}P(x)=\{x_{1},x_{2},\ldots,x_{n}\},\label{population}\end{equation} where
$n$ is the size of the prescribed population and $x_{i}$($0<i\leq n$) are candidate individuals. In this control scheme, each individual is encoded as a sequence of discrete control field values that are
applied to the spin system over time. The initialized solutions are randomized which ensures diversity within the population, allowing for a broad
exploration space of control sequence.

2. Fitness Evaluation: We analyze the evolution of the spin squeezing parameters in a certain time interval, the values are employed to quantify the
performance of each control sequence with the function: $F(x_i)=R(x_i)-R(x)_{min}$, where $R(x_i)$ is a measure of performance in GA signifying squeezing
degree (final state) after the corresponding control sequence $x_i$ which is explained in detail later and $R(x)_{min}$ is the minimum in the population.
\begin{equation}x_{i}=\left(\Omega_{1t, i},\Omega_{2t, i},\ldots,\Omega_{mt, i}\right)\label{individual}\end{equation} in the population is one encoded individual, where $\Omega_{kt, i} (1\leq k\leq m)$
indicates each control pulse in one-time interval, specifically, from~$kt$~to~$(k+1)t$. We simulate the open collective spin system dynamics under the
influence of the control sequence, corresponding to the individual above.

3. Tendentious Selection and Adaptive Genetic Manipulation: To acquire improved solutions as the iteration processes, we implement an elitism strategy
based on the roulette wheel selection mechanism: individuals exhibiting higher fitness values, indicative of superior spin squeezing performance, are
assigned a higher probability of being selected for the subsequent iteration with a smaller rate of elimination. And the probability of an individual
$x_i$ being selected can be written as: $p_i = \frac{F(x_i)}{\sum_{l=1}^mF(x_l)}$, here $m$ is the total number of individual in the population.
This selection mechanism, mimicking natural selection, ensures that desirable traits (i.e., control sequences leading to enhanced squeezing) are
preferentially propagated through the generations.

To prevent premature convergence and promote the exploration of diverse solutions, we introduce adaptive genetic manipulation on selected individuals,
and the `adaptive' is reflected in regulated crossover and mutation rates.

The crossover rate, determining the probability of genetic information exchange between two parent individuals, dynamically adjusts based on their fitness
difference in our strategy. This dynamic crossover rate $c_d$ can be expressed as:
\begin{equation}
c_d=c_s+(1-c_s)\frac{|F(x_i)-F(x_j)|}{f_h},
\label{crossover}
\end{equation}
where $c_s$(0.8 actually) is the minimum mating probability, $1-c_s$ guarantees the normalization of the two weighted terms and $f_h$ (0.5 in the actual
algorithm) serves as a scaling factor representing the characteristic fitness difference within the population. So for two parent individuals $x_i=\left(\Omega_{1t,i},\Omega_{2t,i}...\Omega_{kt,i},...,\Omega_{nt,i}\right)$ and $x_j=\left(\Omega_{1t,j},\Omega_{2t,j},...,\Omega_{kt,j},...,\Omega_{nt,j}\right)$,
their offspring can be:~$x_c=\left(\Omega_{1t,c},\Omega_{2t,c},...,\Omega_{kt,c},...,\Omega_{nt,c}\right)$
where each $\Omega_{kt,c}(1\leq k\leq n)$ ($\Omega_{kt,c}$ describes the specific control pulse as mentioned in the part of Fitness Evaluation) is acquired
with $\Omega_{kt,c}=\begin{cases}\Omega_{kt,j}, if~U>c_d\\\Omega_{kt,i}, if~U\leq c_d\end{cases}$ and $U$ is a uniform-distributed random number within (0,1).
This formulation ensures that a larger disparity in fitness between parents results in a higher crossover probability, encouraging wider exploration of the
potential excellent offsprings.

Conversely, the mutation rate in this task, which governs the introduction of random self-variations in individuals, is progressively reduced with iteration.
This dynamic mutation rate $m_d$ can be formulated as:
\begin{equation}
m_d=m_s\left(1- \frac{g_t}{g}\right),(g_{t}=1,2,3,...,g),
\label{mutation}
\end{equation}
where $m_s$(0.2 in practical situation) is the maximum mutation rate at the initial generation ($g_t=1$), $g_t$ is the current generation number, and $g$ is
the total number of generations. This dynamic setting balances exploration in early generations, where diversity is crucial, with the maintenance of the
achieved optimization effect as the algorithm converges towards an optimum. Further, maintenance signifies there is less and less probability of variance in
population, that is, tendentiously exploiting existing solutions to merge.

The algorithm iteratively repeats these steps with a predefined number of iterations, progressively updates the population and searching for control field
sequences that decline the spin squeezing parameter furthest.

\section{Collective Spin Model}
\label{CSM}
To verify the effectiveness of our proposed strategy, we concentrate on an ensemble of $N$ identical spins, which can be indicated with pseudo spin components $\hat{J}_{\alpha}=\frac{1}{2}\sum^{N}_{k=1}\hat{\sigma}_{\alpha}^{(k)}$, $(\alpha=x, y, z)$, where $\hat{\sigma}_{\alpha}^{(k)}$ is the Pauli operator for the
$k$-th spin(qubit)~\cite{40PRA62211}. For the symmetric scenario where the operations done on the ensemble have identical impact on all the qubits, $\hat{J}_x$,
 $\hat{J}_y$, $\hat{J}_z$ fulfill the relationship: $[\hat{J}_{\alpha},\hat{J}_{\beta}]=i\hbar\epsilon_{\alpha\beta\gamma}\hat{J}_{\gamma}$,
where $\epsilon_{\alpha\beta\gamma}$ is the L\'evi-Civit\`a symbol. The \textcolor{Goldenrod}{Lipkin-
Meshkov-Glick type} Hamiltonian can be written as:
\begin{equation}
 \hat{H}/\hbar=\kappa \hat{J}_{z}^{2}+\Omega_x(t)\hat{J}_{x},
 \label{HC}
\end{equation}
here the first term is the nonlinear interaction in the one-axis twisting (OAT)-type spin squeezing~\cite{12PRA475138} which can provide the resource for
quantum-enhanced metrology~\cite{12PRA475138,2RMP90035005}. $\kappa$ is the atomic interaction strength which is assumed as the unit ($\kappa = 1$) in this
work. It experimentally depends on the scattering lengths between the particles and the condensate density ~\cite{18Nature4641170} and $\Omega_x(t)$ is
the strength of the transverse external field.

\textcolor{blue}{While our work focuses on optimizing the time-dependent linear coupling term $\Omega_x(t)\hat{J}_{x}$, it is noteworthy that the time-dependent one-twisting term $\hat{J}_{z}^{2}$ has also been explored within the two-state Bose-Hubbard model that maps to the Lipkin-Meshkov-Glick type Hamiltonian $\hat{H}_\mathrm{IBJJ}/\hbar=U(t)\hat{J}_z^2-\Omega \hat{J}_x$ via the Schwinger-boson formalism~\cite{prap20054038}. $\Omega$ is the constant strength of Rabi coupling and $U(t)$ is the variable two-body interaction strength owing to the presence of Feshbach resonance~\cite{prl105204101} which can potentially be designed through GA, similar to the optimization of $\Omega_x(t)$ in Hamiltonian~\eqref{HC} discussed later.}
\textcolor{ForestGreen}{When rewritten in terms of Pauli
operators of individual spin-1/2 particles – the OAT term
in the Hamiltonian~\eqref{HC} becomes equivalent to all-to-all Ising coupling
between qubits, while the full Hamiltonian~\eqref{HC} describes a system of all-to-all
Ising-coupled qubits in a time-dependent transverse control field, and such systems have recently attracted considerable attention in the context of
engineering maximally-entangled multiqubit states~\cite{pra106052613,pra108012608}.}
\textcolor{Goldenrod}{Besides, the Hamiltonian~\eqref{HC}, which also describes
systems such as bosonic Josephson junctions~\cite{pra110022610}, corresponds to what became
known as the “twist-and-turn” type dynamics of spin squeezing. 
The efficient and robust generation of spin-squeezed states have been explored in the scenario of time-dependent linear coupling by using the method of shortcuts of adiabaticity~\cite{pra110022610}.}

The collective spin dynamics can be transformed to its dual bosonic representation through Schwinger's transformation methodology: $\hat{J}_z=\frac{1}{2}(\hat{a}^\dagger\hat{a}-\hat{b}^\dagger\hat{b})$,
$\hat{J}_+=\hat{a}^\dagger\hat{b}$ and $\hat{J}_-=(\hat{J}_+)^\dagger$, where $\hat{a}$($\hat{a}^\dagger$) and $\hat{b}$($\hat{b}^\dagger$) are the two
annihilation (creation) operators of two boson modes. In this perspective, the assignment of one mode to represent the spin-up state and the other to
signify the spin-down state allows $\hat{J}_z$ to encapsulate the disparity in population between the two modes within the framework of a Ramsey interferometer~\cite{41AMQP1981,3PRA46R6797,12PRA475138,2RMP90035005}.
Analogous to a linear beam splitter in interferometry, the coupling Hamiltonian term $\hat{J}_{x}$ effects a rotation of the collective spin by an angle $\theta=\int_{t_{0}}^{t_{0}+\Delta t}\Omega(t)dt$ around the $x$-axis over a time interval $\Delta t$.

In distinction from the utilization of a static control~\cite{14PRA63055601} and periodic control~\cite{42ADP2400056}, the present proposal employs a genetic
algorithm optimizing the external control field $\Omega_x(t)$ as a sequence of rectangular pulses, to prepare nonclassical states.
The pulsed control is operationally analogous to a series of linear beam splitters in an interferometer, manipulating the spin system~\cite{17Nature4641165,18Nature4641170}.
The linear control Hamiltonian's capability to steer the spin system arises from the non-commutability: $[\hat{J}_z^2,\hat{J}_{x}]=i(\hat{J}_{y}\hat{J_z}+\hat{J}_z\hat{J}_{y})\neq0$.

The degree of spin squeezing can be effectively quantified using parameters structured from the expectation values of collective spin operators, as detailed
in seminal works~\cite{1PR50989,2RMP90035005}. 
Achieving variances below the standard quantum limit signifies the onset of spin squeezing, rendering the system
suitable for precision metrology by virtue of amplified sensitivity in the spin components perpendicular to the mean-spin direction.
A crucial parameter for quantifying spin squeezing is the \textcolor{red}{Wineland spin-squeezing parameter (coherent spin squeezing parameter)}, defined as:
\begin{equation}\begin{split}\xi_{\perp}^{2}&=\frac{N \min(\Delta\hat{J}_{\vec{n}_{\perp}}^{2})}{|\langle \hat{J}_o\rangle|^2}=\left(\frac{J}{|\langle\vec{J_o}\rangle|}\right)^2 \frac{4\min\left(\Delta J_{\vec{n}\perp}^2\right)}{N}\\&=\frac{N\left[\left\langle \hat{J}_{\vec{n}_{1}}^{2}+\hat{J}_{\vec{n}_{2}}^{2}\right\rangle -\sqrt{\left\langle \hat{J}_{\vec{n}_{1}}^{2}-\hat{J}_{\vec{n}_{2}}^{2}\right\rangle ^{2}+\langle \left[\hat{J}_{\vec{n}_{1}},\hat{J}_{\vec{n}_{2}}\right]_+\rangle^{2}}\right]}{2|\langle \hat{J}_o\rangle|^2},\end{split}\end{equation}
where \textcolor{red}{$N=2J$ is the total number of spins}, $\hat{J}_{\vec{n}_{i}}$ is the collective spin component along the unit vector $\vec{n}_{i}$ ($i=1,2$), defined as $\hat{J}_{\vec{n}_{i}}$=$(\hat{J}_x,\hat{J}_y,\hat{J}_z)\cdot \vec{n}_{i}$, ($i=1,2$). The specific directions are given by $\vec{n}_{1} =(-\sin{\phi},\cos{\phi},0)$
and $\vec{n}_{2} =(\cos{\theta}\cos{\phi},\cos{\theta}\sin{\phi},-\sin{\theta})$. The collective spin operator along the mean spin direction is defined as $\hat{J}_o$=$(\hat{J}_x,\hat{J}_y,\hat{J}_z)\cdot (\sin \theta \cos\phi,\sin \theta \sin \phi, \cos \theta)$, with $\theta$ and $\phi$ representing the polar and
azimuth angles of the mean spin vector, respectively. These angles are determined by $\theta$ = $\arccos(\frac{\langle \hat{J}_z\rangle}{|{\hat{J}}|})$ and $\phi$ = $\mathrm{sign}(\langle\hat{J}_y\rangle)\arccos(\frac{\langle \hat{J}_{x}\rangle}{|{\hat{J}}|\sin\theta})$, where $|{\hat{J}}|=\sqrt{\langle \hat{J}_{x}\rangle^{2}+\langle \hat{J}_{y}\rangle^{2}+\langle \hat{J}_{z}\rangle^{2}}$ is the magnitude of the mean spin vector \cite{1PR50989,2RMP90035005,43JPB39559}.
The direction $\vec{n}_{\perp}$ corresponding to the minimum spin variance is given by $\vec{n}_{\perp}$=$\vec{n}_{1}\cos\varphi+\vec{n}_{2}\sin\varphi$, where the angle $\varphi$ corresponds to the best direction with the lowest squeezing parameter, satisfying
\begin{equation}
	\varphi=\left\{ \begin{array}{ll}
		\frac{1}{2}\arccos\Big(\frac{-A}{\sqrt{A^{2}+B^{2}}}\Big) & \text{if}~~B\leq0,\\
	\pi-\frac{1}{2}\arccos\Big(\frac{-A}{\sqrt{A^{2}+B^{2}}}\Big) & \text{if}~~B>0.\end{array}\right.\label{minimumang}
\end{equation}
Here, $A\equiv\langle J_{\vec{n}_{1}}^{2}-J_{\vec{n}_{2}}^{2}\rangle,\text{ \ }\; B\equiv\langle \left[\hat{J}_{\vec{n}_{1}},\hat{J}_{\vec{n}_{2}}\right]_+\rangle$
are defined for brevity.
\textcolor{red}{The parameter $\xi_{\perp}^{2}$ quantifies the achievable phase sensitivity $\Delta\theta = \xi_{\perp} / \sqrt{N}$, directly linking spin squeezing to metrological enhancement~\cite{16Nature40963}.}

In this control scheme, we employ the following definition
\begin{equation}
\xi_{Z}^2=\frac{4\Delta \hat{J}_{z}^2}{N}
\label{Xis}
\end{equation}
\textcolor{red}{acquired from Kitagawa-Ueda spin-squeezing parameter (number-squeezing parameter)}, where $\Delta \hat{J}_{z}^2$ = $\langle \hat{J}_z^2\rangle-\langle \hat{J}_z\rangle^2 = \min(\Delta\hat{J}_{\vec{n}_{\perp}}^{2})$ ~\cite{1PR50989}. \textcolor{red}{This definition is valid when the mean spin lies in the equatorial plane and the direction of minimal variance is fixed to the $z$-axis.} \textcolor{red}{
While our optimization targets the Kitagawa-Ueda parameter $\xi_Z^2$ for its generality, the resulting states also exhibit a small $\xi_{\perp}^2$, making them valuable for quantum-enhanced measurements.}

For the GA optimization scheme, the performance function is calculated by:
\begin{equation}
\begin{split}
R_{x_i}&=P_{final}R_{final}+P_{process}\langle R_{process}\rangle,\\ &=P_{final}R(\Omega_{mt,i})+P_{process}\frac{\sum^{m-1}_{q=1}R(\Omega_{qt,i})}{m-1},
\end{split}
\label{rewardall}
\end{equation}
where \begin{equation}R_{final}=\frac{1}{\xi_{Z, final}^2}=\frac{N}{4\Delta \hat{J}_{z, final}^2}\label{reward}\end{equation} is the performance score
corresponding to the final state after the application of the entire control sequence, while $\langle R_{process}\rangle $ is the mean value of the performance
scores corresponding to the sampling points(after each pulse is applied) in the control process calculated with the same formula (Eq.~\ref{reward}). $P_{final}$
and $P_{process}$ are two constants(0.8 and 0.2 in simulation) for the entire training session, representing the contribution weights of the degree of final state
squeezing and the average degree of process states to the evaluation of the quality of one control sequence. $R_{x_i}$ corresponding to each individual is an
important basis for calculating individual fitness at the population level, as the evaluated fitness value is defined with the difference between individual
performance value and the smallest performance value in the population. Subsequently, compared to using the reverse of $\xi_{\perp}^{2}$ with variable $\theta$
and $\phi$, the choice of $\xi_{Z}^2$ possesses a more direct physical interpretation since $\hat{J}_z=\frac12\left(\hat{a}^\dagger \hat{a}-\hat{b}^\dagger \hat{b}\right)=\frac12\left(\hat{N_a}-\hat{N_b}\right)$ indicates the population imbalance between the two modes as well as the phase difference between two detectors
within the Ramsey interferometer~\cite{3PRA46R6797,12PRA475138,2RMP90035005}, while $\hat{N_a}$ and $\hat{N_b}$  signify the atom number of two modes.

\begin{figure*}[htbp]
	\includegraphics*[width=18cm]{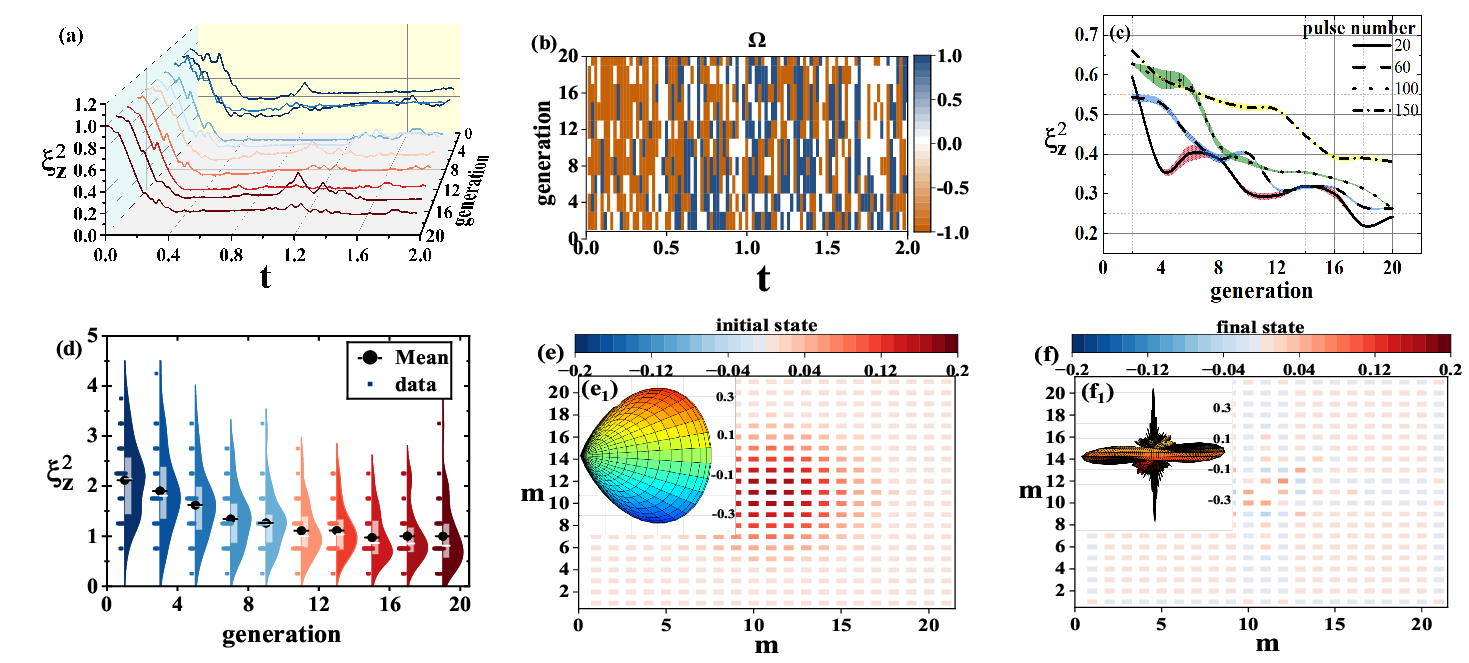}
	\caption{(a) Evolution of the spin squeezing parameter while 100 rectangular pulses are taken across each evolution time interval $[0,2]$ and a total of
20 consecutive training rounds are shown with 10 iteration curves (one curve is displayed every two generations). (b) Schematic representation of the three-level ($\Omega(t)
=1,0,-1$) square control pulses used in (a). (c) The evolution of average spin squeezing parameter with different pulse
numbers and corresponding variance of squeezed state at $t=2$ during iterations. (d) Evolutionary trend of the statistical distribution, average, and median
of final-state squeezing parameter values for individuals within uniform-sampled 10 generations, and each `violin' pattern corresponds to one generation as
well as to one population. Solid black balls represent the mean, and white bars represent the median. (e)The real part of the density matrix of the initial
CSS and($\rm e_1$) is the Wigner representation of it. (f)The real part of the matrix of the spin squeezed state at $t=2$ and ($\rm f_1$) is its corresponding
Wigner representation. $\gamma/\kappa = 0.001$ and $\gamma_z/\kappa = 0.001$ in these results.}
    \label{DoubleCFxi}
\end{figure*}

\section{Procedure to prepare spin-squeezed states using Genetic Algorithm}
\label{PSSS}
The procedure for generating spin-squeezed states encompasses two principal stages: initially, the establishment of a spin coherent state, followed by the
application of a genetically optimized control sequence, to achieve the desired spin-squeezing.
\subsection{Initial coherent spin state}
A collection of $N$ identical two-level atoms, each aligned in the same direction, is aptly represented by SU(2) coherent spin state (CSS), which can be
articulated as the projection of a coherent state onto Dicke states $|j,m\rangle$, reads
\begin{equation}
\begin{split}
|\eta\rangle&\equiv|\theta,\phi\rangle\\&=(1+|\eta|^2)^{-j}\sum_{m=-j}^j\binom{2j}{j+m}^{1/2}\eta^{j+m}|j,m\rangle,\\
&=\exp\left(\eta \hat{J}_+\right)\exp\left[\ln\left(1+|\eta|^2\right)\hat{J}_z\right]\exp\left(-\eta^*\hat{J}_-\right)|j,j\rangle,
\end{split}
\end{equation}
where$\quad\eta=-\tan\frac{\theta}{2}\exp(-i\phi)\in\mathbb{C}$, $|j,m\rangle$ are the eigenstates of $\hat{J}_{x}$ with eigenvalue $m$, satisfying the
equations: $\hat{J}^2|j,m\rangle$=$j(j+1)\hbar^2|j,m\rangle$ and $\hat{J}_z|j,m\rangle$=$m\hbar|j,m\rangle$ ($\hbar$=1 in numerical calculations); $|j,j\rangle\equiv\bigotimes_{l=1}^N|0\rangle_l$  represents a state with all the spins polarized in the z-direction. This overcomplete state is most similar
to the classical one with $\theta$ and $\phi$ being the azimuth angles for longitude and latitude, respectively. The quantum state can be generated by the effect
of rotation operator $\hat{R}(\theta,\phi)=e^{-i\theta\hat{J}_{\vec{n}}}=e^{i\theta(\hat{J}_z\sin\phi-\hat{J}_y\cos\phi)}$\cite{1PR50989}, where $\vec{n}=(-\sin\phi,\cos\phi,0)$ on the eigenstate of $\hat{J}_z$ with the applicable Husimi Q-function:\begin{equation}Q(\theta,\phi)=\langle\theta,\phi| \rho |\theta,\phi\rangle\end{equation} visualizing the probability density of quantum state having coordinates $\theta$ and $\phi$ in the phase space, $\rho$ is the density matrix of the collective spin system in the reservoir.
It can also be represented by the Wigner distribution which is calculated in the collective model as
$W_{jm}(\theta,\varphi)=\sum_{k = 0}^{2j}Y_{k0}(\theta,\varphi)(-1)^{j - m}\sqrt{2k+1}\times\left[\begin{array}{ccc}j&k&j\\-m&0&m\end{array}\right]^*$,
where $\left[\begin{array}{ccc}j&k&j\\-m&0&m\end{array}\right]^*$ is the Wigner 3$j$ symbol~\cite{PRA242889} and $Y_{k0}$ is the spherical harmonic. Furthermore, the
CSS can be prepared by applying $\pi/2$ pulses to a BEC with $N$ atoms in the internal ground state~\cite{16Nature40963,17Nature4641165,18Nature4641170},where $\langle \hat{J}_{x}\rangle=N/2=J$ and $\langle \hat{J}_{y}\rangle=\langle \hat{J}_{z}\rangle=0$ and the internal wave function $|\Psi\rangle$ of the BEC is thus written by: $|\Psi(t=0)\rangle=\frac1{\sqrt{2^NN!}}(a^\dagger+b^\dagger)^N|\text{vac}\rangle,$ where~$|\text{vac}\rangle$ is the vacuum state.
Starting with a CSS aligned along the $x$-axis and exhibiting isotropic fluctuations in its spin components, the application of $\hat{J}^2_z$ induces a shear,
transforming the state into a spin-squeezed state with reduced variance in $z$ direction. This results in a state that surpasses the standard quantum limit,
thereby enhancing metrological measurement sensitivity along the direction of squeezing~\cite{44science1104149}. The direction of squeezing is dynamically
adjusted by the action of $\Omega(t)\hat{J}_x$, which is initially randomized and subsequently refined through optimization by the GA as previously described.

\subsection{Prepare spin-squeezed states by GA-optimized pulses}

Preserving the integrity of quantum states against the detrimental effects of decoherence, induced by inevitable interactions with the environment, is a formidable
challenge in quantum technologies. Effective quantum control strategies are thus crucial for mitigating decoherence and safeguarding the delicate quantum resources
that underpin coherent quantum operations. This study investigates the impact of two prominent decoherence channels, namely, superradiant damping and dephasing, both
notorious for their capacity to erode quantum coherence and compromise quantum information processing.
The temporal evolution of the collective spin dynamics is rigorously described by the Lindblad master equation, formulated as:
\begin{eqnarray}
\dot{\rho}=-i[\hat{H},\rho]+\gamma(n_{th}+1)\mathcal{L}_{\hat{J}_{-}}\rho+\gamma n_{th}\mathcal{L}_{\hat{J}_{+}}\rho+\gamma_{z}\mathcal{L}_{\hat{J}_{z}}\rho,
\label{eqmaster}
\end{eqnarray}
where $\mathcal{L}{\hat{X}}\rho = 2 \hat{X}^{\dag} \rho \hat{X} - \hat{X} \hat{X}^{\dag} \rho - \rho \hat{X} \hat{X}^{\dag}$, $\rho$ is the density matrix for the controlled system. Here, $\gamma$ denotes the decay rate,
$\gamma_z$ represents the dephasing rate, and $n_{th}$ signifies the average thermal photon number.
Unlike conventional quantum Lyapunov control methods, which focus on minimizing the distance between eigenstates, specifically, increasing the inner product between
current state and target state~\cite{45PLA425127874}, our approach leverages a GA to optimize the control sequence $\Omega(t)$ within the Hamiltonian Eq.~\ref{HC},
thereby guiding the system's evolution under the Lindblad master equation.

Our approach employs a GA to identify optimal sequences of control pulses for driving a spin system toward a desired spin-squeezed state. The algorithm explores a
population of candidate control sequences, where each individual within this population represents a sequence of square pulses. We quantify the effectiveness of each
control sequence (Eq.~\ref{rewardall}) by simulating the evolution of the spin system under its influence, according to the master equation Eq.~\ref{eqmaster}. The
degree of spin squeezing achieved at the end of the evolution serves as a main fitness measure (Eq.~\ref{rewardall}), with higher fitness values assigned to sequences
yielding stronger squeezing (Eq.~\ref{reward}).

This iterative optimization process proceeds as follows. Each control sequence is encoded and then evaluated by quantifying the resulting spin squeezing after the
simulation of the system's dynamics. Based on these fitness evaluations, a selection process preferentially propagates higher-fitness sequences to subsequent generations, mimicking natural selection. To further enhance exploration and prevent convergence to local optima, crossover and mutation operators are introduced, introducing
randomness into the offspring generation. Respectively, crossover facilitates the combination of beneficial traits from different parent sequences, while mutation stochastically explores new regions of the control parameter space.

It is crucial to emphasize that while the optimization process utilizes a closed-loop simulation, the resulting control strategy is implemented in an open-loop manner.
This distinction is critical for practical applications. During the optimization, once the optimal control sequence, such as the $\Omega_{x}(t)$ defined in Eq.~\ref{HC},
is identified, it is applied in a single, uninterrupted open-loop control process. This avoids wavefunction collapse associated with continuous measurements during the
control implementation, preserving the coherence necessary for achieving the desired spin-squeezed state.

\section{Control Results}\label{ConRes}
The control protocol, illustrated in FIG.~\ref{DoubleCFxi}, divides the time interval [0, 2] into a variable number of segments. Square control pulses are applied at the boundaries between the adjacent time segments and sustained until the subsequent boundary.
The control pulse sequences are generated randomly in groups at the beginning and optimized and renewed by GA.
The GA optimizes these pulse sequences over 20 generations and every 2 generations of the evolution for the squeezing parameter are illustrated in FIG.~\ref{DoubleCFxi} (a).
It can be seen that, at the first stage of the training, the GA has already found the effective control sequence, resulting in the downward trend of $\xi_Z^2$.
As the training proceeds, even after a small number of iterations, the GA can find a sequence of square pulses inducing an optimal performance of squeezing, however, it does not show a further improvement in subsequent iterations before the end.
FIG.~\ref{DoubleCFxi} (b)illustrates the timing of the corresponding square wave control pulses.
In FIG.~\ref{DoubleCFxi} (c), when the frequency of pulse application is higher, the squeezing effect of the final state of each generation changes more gradually with the iteration, but the overall squeezing effect is poor.
Corresponding to~FIG.~\ref{DoubleCFxi} (d), the statistical distribution of final-state squeezing parameters for individuals within the evolving population reveals a clear trend: as the optimization control progresses, the proportion of relatively advantageous individuals (characterized by $\xi_Z^2<1$) steadily increases.
These results validate the GA optimization strategy for enhancing control pulse performance.

It's worth mentioning that the distribution of squeezing parameter adheres to the principles of Kernel Density Estimation (KDE), a non-parametric statistical method employed to estimate the probability density function (PDF) of a dataset and derive a smooth representation of the underlying distribution, which offers a robust alternative to parametric methods, as it avoids imposing assumptions about the specific form of the distribution (e.g., Gaussian, exponential). The kernel density estimator's formula is given by $\widehat{f}_h(x)=\frac{1}{nh}\sum_{i=1}^nK\biggl(\frac{x-x_i}{h}\biggr) ,$ where $x_{1},x_{2},...,x_{n}$ are acquired samples from an unacquainted distribution, $n$ is the sample size, $h$ is the bandwidth and $K(\cdot)$ is the kernel function~\cite{oup1997}. Instead, KDE leverages $K(\cdot)$ to smoothly distribute the probability mass associated with each data point, resulting in a continuous and differentiable estimate of the PDF.
Consequently, both the average(black dots with black horizontal bars) and median(white horizontal bar) squeezing coefficients exhibit a decline. However, due to the intentional preservation of a degree of exploratory freedom (mutation rate~$m_{d}\geq0$) and the finite duration of the iterative process, it is not possible to achieve effective control for all individuals within the population during the tested number of iterations.

Furthermore, FIG.~\ref{DoubleCFxi} ($\rm e_1$) and ($\rm f_1$) utilize the Wigner function to visualize the initial coherent
spin state and final spin-squeezed state at the time $t=2$.  The irregular folds in Wigner-like function is relevant to non-classical effect which leads to interesting interference effects~\cite{48pra86062117}.
This provides insight into the evolution of the quantum state
under control. To show the squeezing process more
vividly, a movie of the squeezing process is shown by the Husimi function in~\cite{49Movie}.
\begin{figure}
	\includegraphics[width=8.6cm]{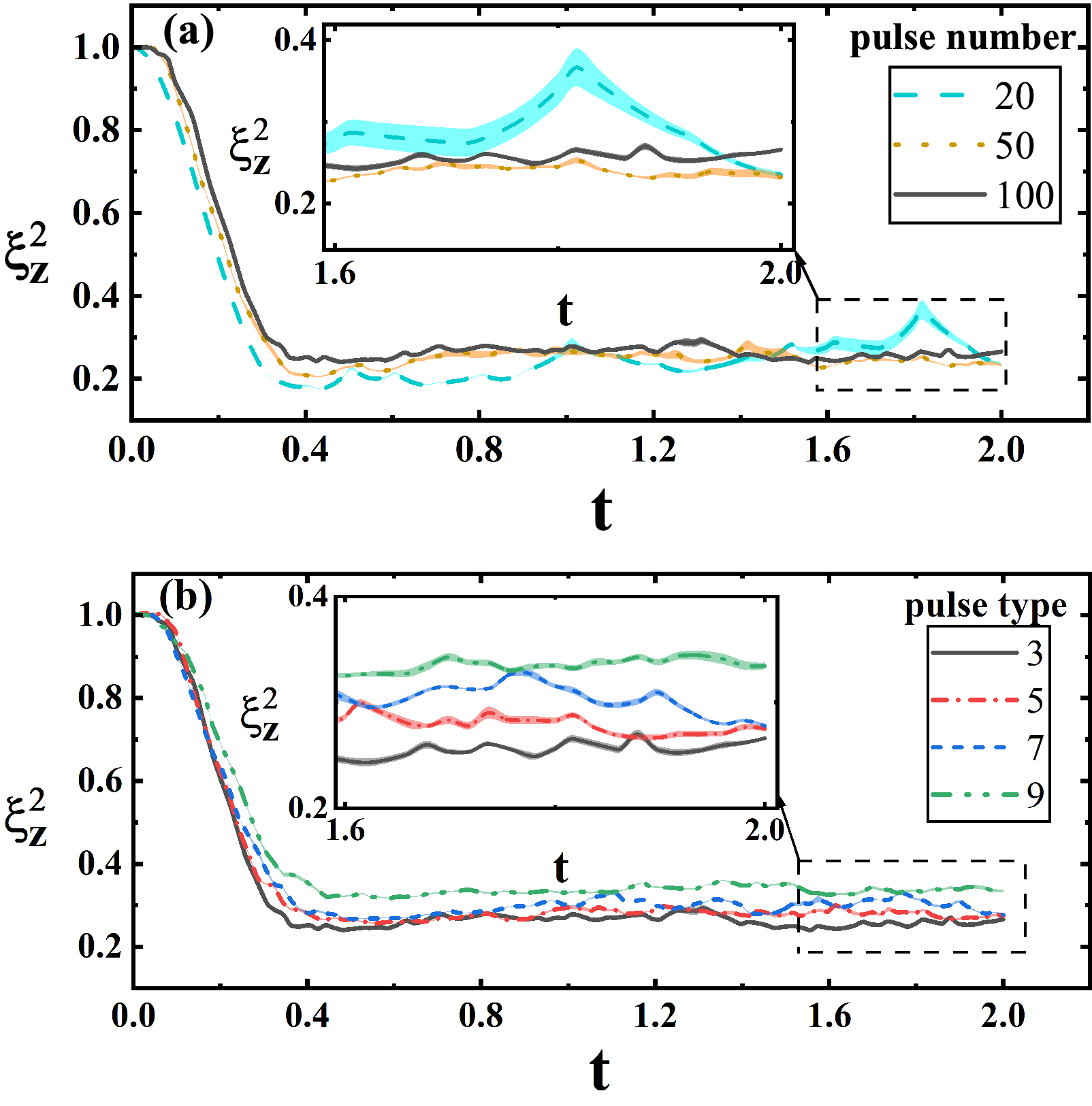}
	\caption{(a) Mean evolution of the last iteration sample of $\xi_{Z}^2$ in 5 repetitions every 30 generations.
The shaded regions indicate the variance for different frequencies of applying rectangular pulses. Control time interval $[0,2]$ is divided into the
number of specific segments at which the square pulses
with three levels($\Omega(t)=1,0,-1$) are applied.
Samples were all obtained from the final generation results of each training session.
(b) Evolution of the spin squeezing parameter for different numbers of pulse gears when 100 total pulses are applied, and the correspondence between the number of pulses and their intensity is as follows:
$\{\Omega(t)\}$=
$\{(1, 0, -1)|~actions=3\}$,
$\{(1, 0.5, 0, -0.5, -1)|~actions=5\}$,
$\{(1, 0.67, 0.33, 0, -0.33, -0.67, -1)|~actions=7\}$,
$\{(1, 0.75, 0.5, 0.25, 0, -0.25, -0.5, -0.75, -1)|~actions=9\}$.
}
\label{DoubleCFeacc}
\end{figure}
We investigate the impact of pulse frequency on control effectiveness by discretizing the time interval [0, 2] into different numbers of segments (more segments corresponds to higher frequency).
FIG.~\ref{DoubleCFeacc}(a)shows that higher frequencies lead to lower variance in the squeezing parameter but a slower decrease of mean value at early steps.
This reveals a trade-off between control performance and experimental complexity:
if we pursue a more stable control effect on the squeezing parameter, more pulses are needed in a certain time period, which is more demanding for the experimental device.
Because the genetic algorithm is essentially the optimization of the sequence, it is not limited by the sequence length, and the sequence length in this task only depends on the number of controls (the number of time segments) in the total time scale, which can theoretically optimize the high-frequency-varying control sequence.

We also investigate the influence of the number of pulse gears on control performance.  As shown in FIG.~\ref{DoubleCFeacc}(b), there is an obvious advantage
for fewer control types with the same maximum control amplitude in
this control, which conclusion is different from the previous conclusion of using reinforcement learning for control.
The reason may lie in the following: with the number $p$ of control pulses remaining constant in total time $[0,2]$, when the number of control types increases from $q$ to $q+d$, the search space for the optimal control sequence explored by GA expands by $\left(\frac{q+d}{q}\right)^p$. Consequently, within the same limited number of iterations, the probability of finding an optimal control sequence becomes smaller.

\begin{figure}
	\includegraphics*[width=8.6cm]{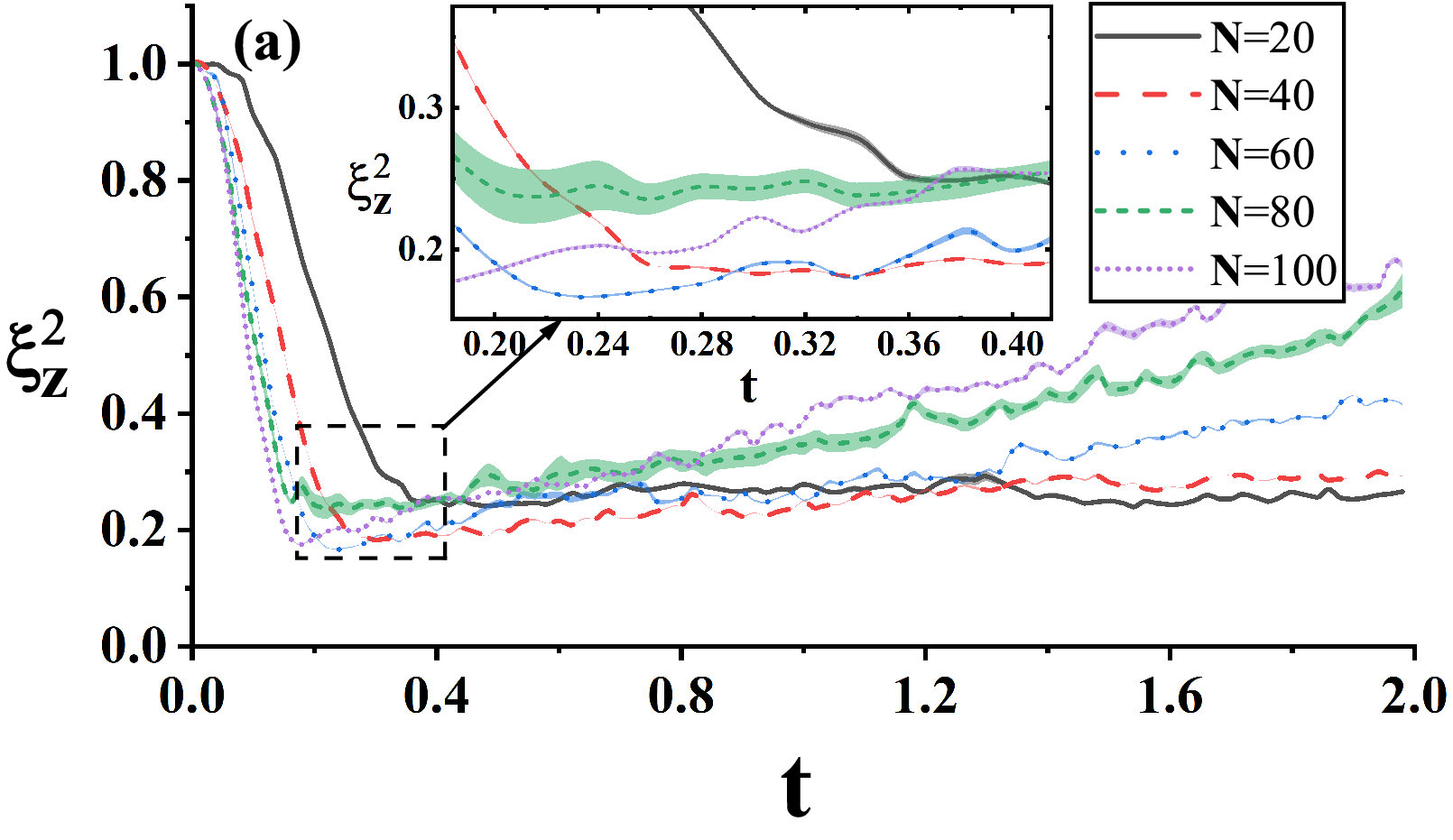}
	\caption{Evolutions for the last iteration of $\xi_{Z}^2$ every 30 generations for different sizes of the
collective spin system $N = 2J$ under three-type ($\Omega(t)=1,0,-1$)
control. The error zone corresponding to variance is calculated by five repetitions. The samples are picked in the same manner as those in FIG.~\ref{DoubleCFeacc} (a) and number of segments is 100.}
\label{RLlearning3size}
\end{figure}

The generalizability of our approach is investigated by applying it to collective spin models with different total spin number ($N=2J$).
FIG~\ref{RLlearning3size} presents the control results for different $N$s.
The results show that the larger the $N$, the faster the compression parameter
drops at the beginning, however, the corresponding squeezing parameter rebounds
more strongly at the final time cut-off.
This observation indicates that a lessened ensemble of spins benefits enhancing the
precision of quantum metrology. 

\begin{figure}
\includegraphics*[width=8.6cm]{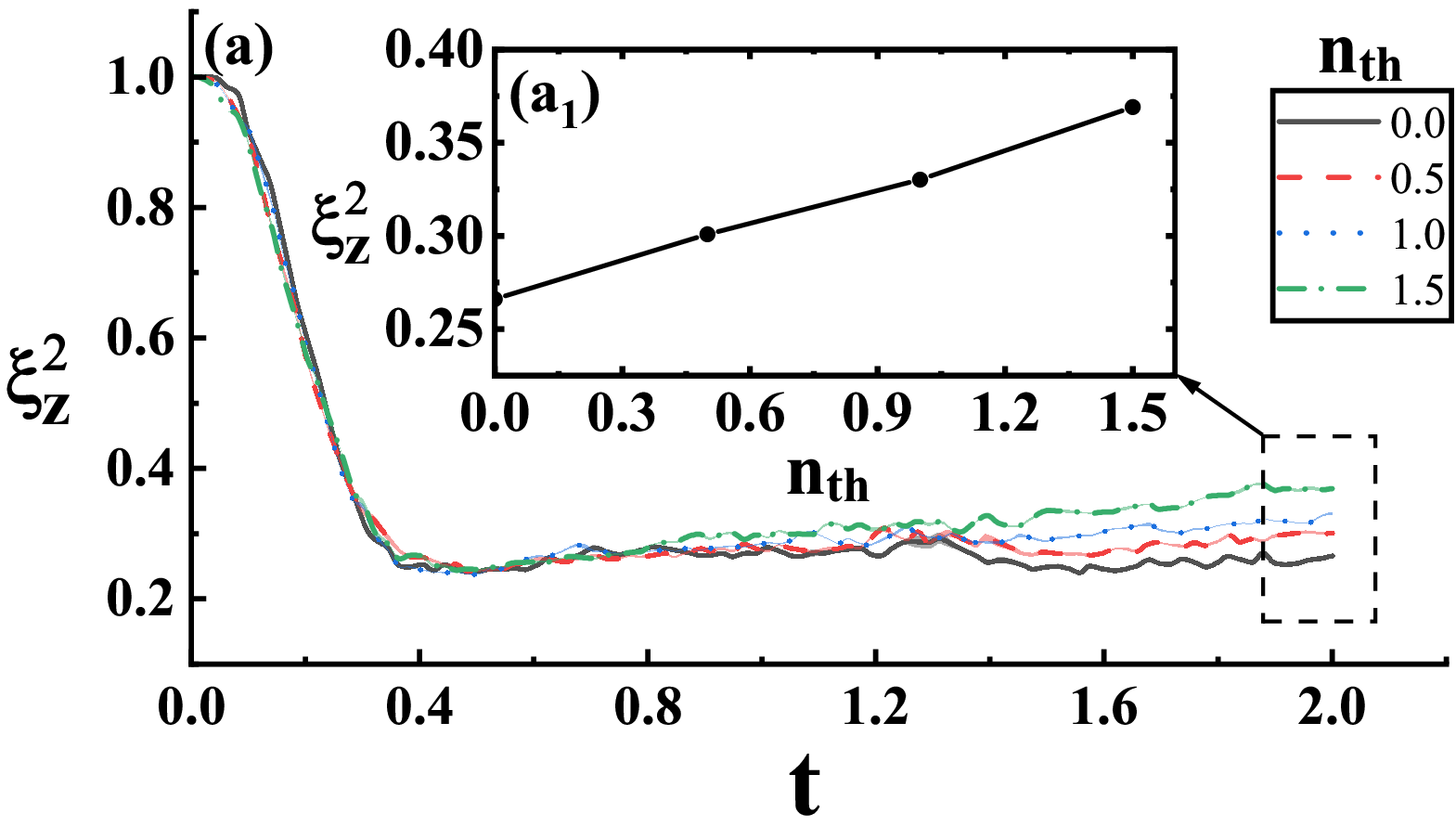}
\caption{The average evolutions of the last iteration of $\xi_{Z}^2$ in 5 repetitions
every 30 generations with variance for different thermal excitations: the average
number of photons for a mode with frequency $\omega$ in the
reservoir. The samples are picked in the same manner as those
shown in FIG.~\ref{RLlearning3size}. The subgraph shows the
squeezing parameter versus average thermal excitation at $t=2$.}
\label{RLlearning3temp}
\end{figure}

The environmental temperature was assumed
to be zero in the previous results. To investigate the robustness of the proposed control
scheme, it is essential to examine how temperature affects the control outcomes. Since the temperature is positively correlated
with the average number of thermal excitations in the reservoir,
denoted by $n_{th}=\frac{1}{\exp^{\hbar\omega/k_BT}-1}$~\cite{50springer10978}, it indicates the strength of the decoherence. Here, $T$ is the temperature of the reservoir.
As shown in FIG.~\ref{RLlearning3temp}, a gradual increase in thermal excitations reduces the effectiveness of the control strategy.

\textcolor{Goldenrod}{According to FIG.~\ref{DoubleCFeacc}, \ref{RLlearning3size} and \ref{RLlearning3temp}, the timescale of achieving the maximal statistical averaging spin squeezing is unaffected by the number of pulses, the pulse type, or the number of thermal excitations $n_{th}$, consistently occurring around t=0.4. The sole influential variable is the particle number $N$, and the timescale is shortened markedly as the $N$ grows.}

Substantially, in the system we analyze, the applied coherent pulses interfere with the energy dissipation that would otherwise drive the quantum system to its ground state. This competition establishes a dynamic equilibrium, enabling stable maintenance of the squeezing parameter.

\begin{figure*}
\includegraphics*[width=18cm]{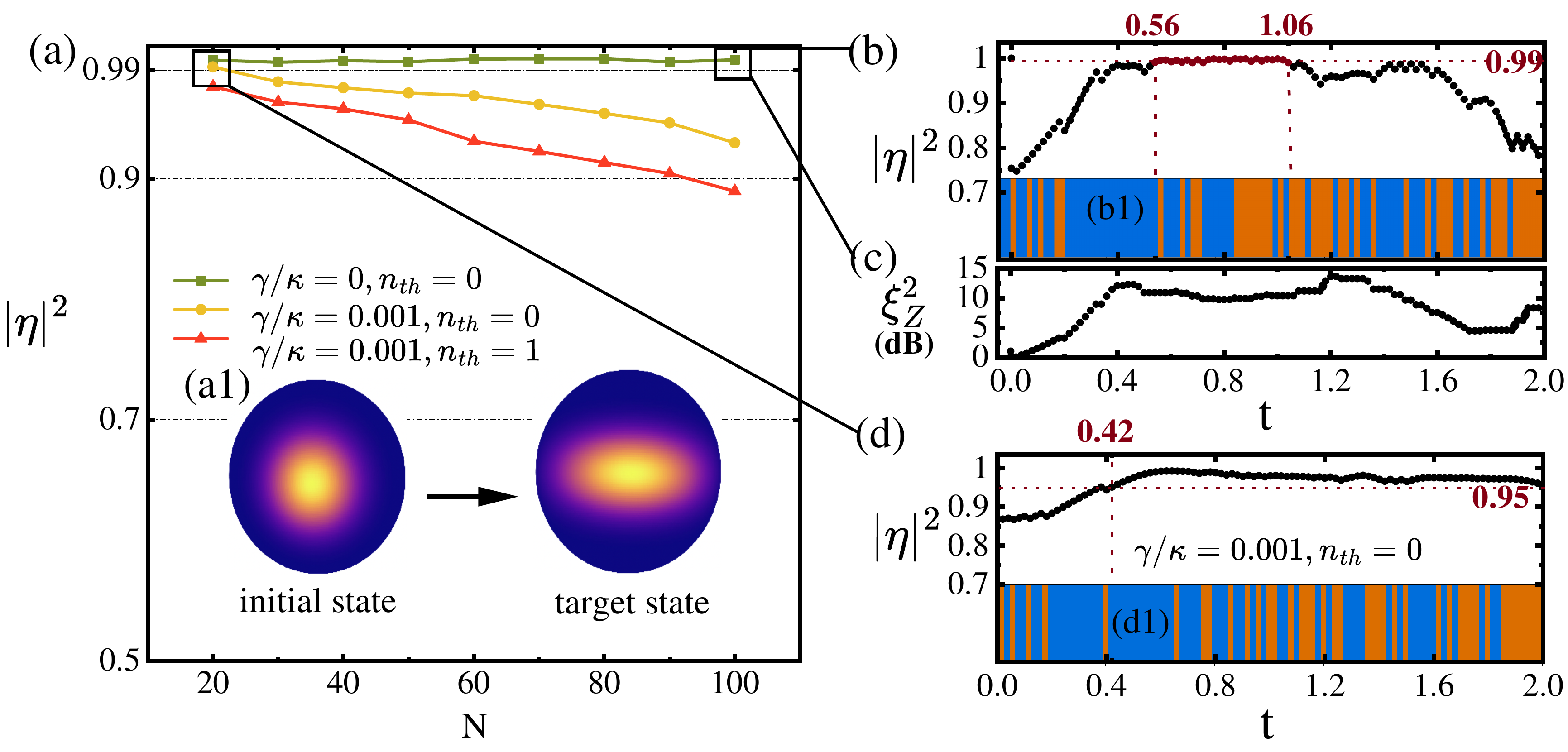}
\caption{Performance and dynamics of the GA-optimized state preparation protocol. \textbf{(a)} Maximum fidelity $|\eta|^2$ achieved by the GA versus the particle number $N$, using the fidelity for the ground state of Hamiltonian~(\ref{HC}) when $\Omega_x(t)/\kappa = 1$ as the reward function. Different colors correspond to different dissipation rates $\gamma/\kappa$ and thermal particle numbers $n_{\text{th}}$.
\textbf{(b)} The time evolution of the fidelity corresponding to $N=100$ in panel (a).
\textbf{(b1)} The control pulse sequence for the $\hat{J}_x$ term used in (b).
\textbf{(c)} The squeezing parameter curve $\underset{(\mathbf{dB)}}{\operatorname*{\operatorname*{\operatorname*{\operatorname*{\operatorname*{\xi_Z^2}}}}}}$ corresponding to the fidelity curve in (b). 
\textbf{(d)} The time evolution of the fidelity corresponding to $N=20, \gamma/\kappa = 0.001, n_{th} = 0$ in panel (a).
\textbf{(d1)} The control pulse sequence for the $\hat{J}_x$ term used in (d).
The results are based on the algorithm parameters: a minimum crossover probability of $c_s = 0.7$ defined in~(\ref{crossover}), a maximum mutation rate of $m_s = 0.3$ in~(\ref{mutation}), a sequence of $m = 100$ (defined in (\ref{individual})) control pulses, and a population size of $n = 100$ defined in (\ref{population}).}
\label{fidelity}
\end{figure*}

\begin{figure}
\includegraphics*[width=8.6cm]{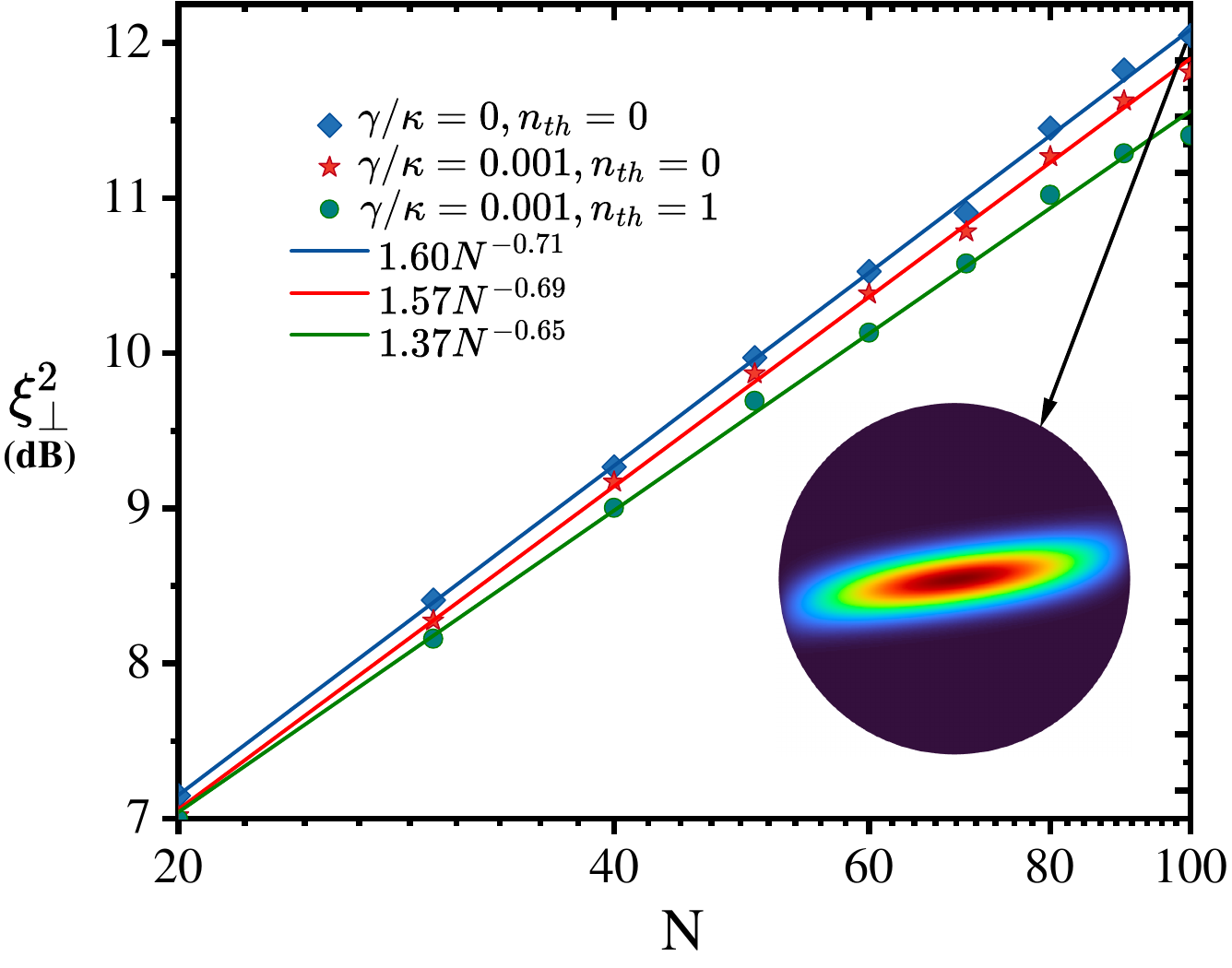}
\caption{Scaling of GA-optimized spin squeezing versus particle number $N$.  Data points correspond to different environmental conditions: an ideal system ($\gamma/\kappa=0, n_{th}=0$, blue diamonds), a system with dissipation ($\gamma/\kappa=0.001, n_{th}=0$, red stars), and a system with dissipation and thermal noise ($\gamma/\kappa=0.001, n_{th}=1$, green circles). The colored solid lines represent power-law fits to the numerical data, with the corresponding scaling laws detailed in the legend. }
\label{scale}
\end{figure}

\textcolor{red}{Inspired by recent advances in VQCs for state preparation~\cite{nrp3625, prl123260505, prap17064050, prr5043285, prr7l022072} and exploration of controlling different Hamiltonian items~\cite{prap20054038,prl105204101,pra106052613,pra108012608,pra110022610}, we employ a control paradigm based on alternating the native one-axis twisting interaction ($\propto$$\hat{J}_z^2$) with a transverse control field ($\propto$$\hat{J}_x$). The genetic algorithm optimizes the temporal arrangement of the control pulse sequence to prepare a specific spin-squeezed target state, with the performance summarized in FIG.~\ref{fidelity}.}

\textcolor{red}{As shown in FIG.~\ref{fidelity}(a), for a dissipationless system ($\gamma = 0, n_{\text{th}} = 0$), our protocol robustly prepares the target state with a fidelity $|\eta|^2$ exceeding 0.99 for systems ranging from $N=20$ to $100$. Besides, this GA-based strategy is resilient to environmental noise; even in the presence of dissipation, the fidelity mainly remains above 0.9. This level of performance is highly competitive with, and not significantly different from, that achieved by emerging VQC approaches~\cite{prap17064050,prr7l022072}.}

\textcolor{red}{Furthermore, GA-based time-continuous control offers a significant advantage over discrete, gate-based methods. As exemplified by the $N=100$ dissipationless case in FIG.~\ref{fidelity}(b), the GA discovers a control solution that not only reaches $|\eta|^2 > 0.99$ but also sustains the high-fidelity state for a prolonged time window from $t=0.56$ to $t=1.06$. 
Even with dissipation $\gamma/\kappa=0.001$, as FIG.~\ref{fidelity}(d) shows, high-fidelity state can also be generated stably for a long time.
Consequently, the corresponding spin squeezing parameter $\underset{(\mathbf{dB)}}{\operatorname*{\operatorname*{\operatorname*{\operatorname*{\operatorname*{\xi_Z^2}}}}}}$ is maintained at a high value, specifically higher than 10 dB throughout this interval, as shown in FIG.~\ref{fidelity}(c), where $\underset{(\mathbf{dB)}}{\operatorname*{\operatorname*{\operatorname*{\operatorname*{\operatorname*{\xi_Z^2}}}}}} = -10 \log_{10}(\xi_Z^2)$. This ability to maintain a desired nonclassical state over a finite duration is a highly sought-after feature for practical metrological applications that require a stable quantum resource during an interrogation period.
Remarkably, the high performance reported is achieved after only 11 training generations, highlighting the algorithm's rapid convergence.}

\textcolor{ForestGreen}{Beyond preparing a specific target state with high fidelity, we further demonstrate the versatility of our GA framework by tasking it to directly optimize for maximum metrological enhancement. For this, the fitness function was redefined to minimize the spin squeezing parameter $\xi^2_\perp$, thereby directly searching for control sequences that generate the most useful states for quantum-enhanced measurements. The FIG.~\ref{scale} shows the minimum achievable squeezing parameter versus the system size $N$ on a log-linear scale, where the squeezing level is expressed in decibels (dB) for clearer visualization of the scaling behavior; larger $\underset{(\mathbf{dB)}}{\operatorname*{\operatorname*{\operatorname*{\operatorname*{\operatorname*{\xi_\perp^2}}}}}}$ correspond to stronger squeezing. For a dissipationless system, the GA discovers control protocols that yield a squeezing level scaling as $N^{-0.71}$, which demonstrates a remarkable and consistent improvement in squeezing with system size, confirming the powerful scalability of our approach in navigating the vast control landscape.}

\textcolor{ForestGreen}{More importantly, the framework showcases its exceptional robustness against environmental noise. As shown in FIG.~\ref{scale}, even in the presence of dissipation ($\gamma/\kappa=0.001$, red pentagrams) and thermal excitations ($n_{th}=1$, green circles), the GA consistently finds control strategies that maintain strong, scalable squeezing, with scaling exponents of $N^{-0.69}$ and $N^{-0.65}$, respectively. This highlights its significant potential for near-term quantum sensing experiments, where environmental noise is often the primary limiting factor, and underscores the algorithm's power to identify near-optimal quantum resources within the classically inaccessible regime.}

\section{Discussions}\label{discussion}

\subsection{Feasible experiments}\label{exp}
Atomic BECs offer a compelling platform for realizing this proposed control scheme, where interactions between atoms can be finely tuned, making them ideal candidates for demonstrating spin squeezing and quantum control.
In these experiments, the condensate ensemble can be effectively described by the Hamiltonian Eq.~\ref{HC}\cite{18Nature4641170,53PNAS1715105115,54nature4647292,55PRL11310}. Specifically, the experimental realization of our scheme can be based on the hyperfine states of atoms in a BEC, which can naturally map onto the spin-up and spin-down states of our model.

\textcolor{blue}{The nonlinear interaction term $\hat{J}_z^2$ , which is the key to generating a spin-squeezed state, can be controlled directly or indirectly by adjusting the normalized density overlap of the two BEC components~\cite{prap20054038}.} This overlap, in turn, can be precisely tuned using Feshbach resonances~\cite{prl105204101,56nature3926672,58RMP821225}, a technique successfully employed in various experiments exploring spinor BEC dynamics \cite{60PRL802027,61NATURE443312}.
Furthermore, the time-dependent control field, represented by the Rabi frequency $\Omega(t)$,  can be readily implemented with $\pi/2$ microwave pulses. These pulses couple the near-resonant two-photon hyperfine states of a $^{87}$Rb or $^{23}$Na BEC confined in an optical lattice~\cite{18Nature4641170,62Science345424,53PNAS1715105115,64PRL811539,54nature4647292,66nature621728}.
Importantly, our proposed scheme requires only minimal modification of existing experimental setups; only a carefully timed sequence of pulses needs to be implemented. This high degree of compatibility with current BEC experiments makes our approach a promising avenue for realizing optimized spin squeezing in the near future.

\subsection{Diversity of application}\label{ContinuousC}
Optimizing control in systems with continuous variables presents a distinct challenge compared to discrete systems, due to the vastness of the control landscape and the complexity of feedback mechanisms.
In such systems, control fields are continuous functions of time, and finding the optimal time-dependent control requires exploring a vast parameter space.
Despite the added difficulty, and even in the presence of noise from measurement feedback, GA has proven effective in tackling continuous-space control problems, such as its viability of continuously adapting the robot controllers~\cite{GAR113131}.
As noted in various fields, GAs can also handle the optimization of multi-dimensional functions efficiently by employing real-valued encodings and specialized operators.
This capability makes GAs highly applicable to continuous quantum control tasks, such as steering the open-ended evolution of quantum states.

Furthermore, the inherent per-particle symmetry of our control scheme within the interferometer naturally extends its applicability to bosonic systems.
In the large-particle-number limit($N$), the collective spin model can be mapped onto a bosonic model via the Holstein-Primakoff transformation:
$\hat{J}_z=N/2 -\hat{a}^\dagger \hat{a}\simeq N/2$,
$\hat{J}_x=\hbar\frac{\sqrt{2J-\hat{a}^\dagger\hat{a}} \hat{a}+\hat{a}^\dagger \sqrt{2J-\hat{a}^\dagger\hat{a}}}{2}$ and $\hat{J}_y=\hbar\frac{\sqrt{2J-\hat{a}^\dagger\hat{a}} \hat{a}-\hat{a}^\dagger \sqrt{2J-\hat{a}^\dagger\hat{a}}}{2i}$
where $\hat{a}$ ($\hat{a}^\dagger$)denotes the bosonic annihilation (creation) operator~\cite{69PR581098}.
This mapping facilitates the direct application of our control strategy to engineer quantum resources in bosonic systems, further showcasing the broad scope of our approach.

\subsection{Replaceability of the genetic algorithm module}\label{Replace}
GA is used here to implement the control scheme, it can be readily replaced with alternative optimization algorithms possessing similar learning capabilities without altering the fundamental structure of our approach.

Several promising candidates exist for substituting the GA, each with its own strengths and limitations. These include Particle Swarm Optimization~\cite{70a19421948}, Ant Colony Optimization(ACO)~\cite{71a134142}, and Firefly Algorithm(FA)~\cite{72lp2010}, among others.
In ACO, the pathfinding process of each `ant' can be viewed as searching for the optimal arrangement of a sequence and the algorithm mimics the foraging behavior of ants to find the best sequence.
In FA, each `firefly' represents a potential sequence solution, and the `brightness' of a firefly represents the quality of the solution.
The algorithm simulates the flashing and mutual attraction behavior of fireflies to find the optimal sequence.
Analogous to our application of GA for optimizing time-varying control fields, all these algorithms applicable to sequence optimization may theoretically be employed to design the arrangement of time-varying control pulses.
The optimal choice of optimization algorithm will likely depend on specific features of the control problem, such as the dimensionality of the search space, the complexity of the fitness landscape, and computational resource constraints.

The broader principle underlying our control scheme is the utilization of a feedback loop to guide the dynamical evolution of a quantum system toward a desired target state. The optimization module, whether it be a GA or an alternative algorithm, serves as a tool to efficiently navigate the control landscape and identify the most effective pathways for achieving this goal. The key advantage of these optimization techniques lies in their ability to leverage existing knowledge of high-performing solutions while simultaneously exploring new possibilities. This balance of exploitation and exploration allows for the discovery of control strategies that may not be readily apparent through intuitive design or brute-force search methods. By framing the control problem as an optimization task, we open the door to a powerful and versatile toolkit for manipulating quantum systems with high fidelity and precision.

\begin{figure}
\includegraphics*[width=8.6cm]{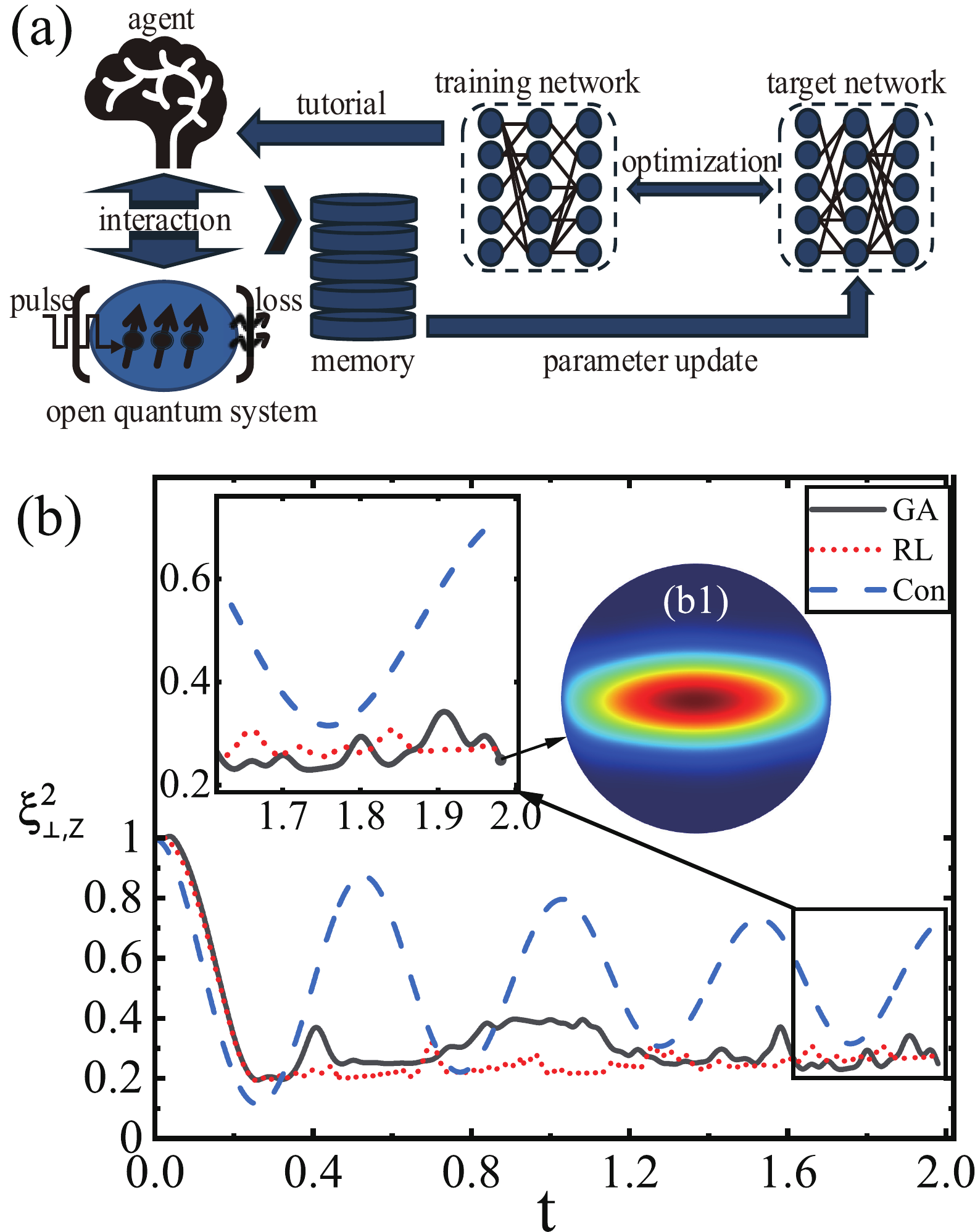}
\caption{Comparative performance of different control strategies. (a) Schematic of the reinforcement learning (RL) control framework. 
The agent interacts with the open quantum system by selecting actions (control pulses). 
The resulting transitions (state, action, reward, next state) are stored in a memory buffer. 
During training, mini-batches of experiences are sampled from the memory to update the `training network'. 
A ‘target network', whose parameters are periodically updated from the training network, is used to provide stable target values for the optimization, mitigating training instabilities.
 (b)The main plot shows the time evolution of the squeezing parameter $\xi_{\perp,Z}^2$ for the adaptive genetic algorithm (GA, solid gray line), reinforcement learning (RL, dotted red line), and a constant control scheme (Con, dashed blue line). Both GA and RL agents were trained for 600 iterations. (b1) The Husimi Q-function of the final spin-squeezed state prepared by the GA protocol, showing a distinct elongated distribution characteristic of strong squeezing. System parameters based: particle number N = 20, dissipation $\gamma/\kappa = 0.001$, pulse size $\Omega(t) \in \{-2, 0, 2\}$.}
\label{RLGA}
\end{figure}

\textcolor{red}{\subsection{Comparative analysis with other strategies}\label{Replace}}

\textcolor{red}{To further benchmark the efficacy of our proposed GA-based strategy, we extend our analysis to include a comparison with both a constant control protocol (Con) and an RL agent. RL represents a mainstream and powerful methodology in classical machine learning, widely recognized for its effectiveness in solving complex control optimization problems~\cite{19Murphy2012,20Sutton2018,21Nature549195,pla539130368}. FIG.~\ref{RLGA} illustrates the time evolution of the spin squeezing parameter $\xi_{\perp,Z}^2$ under these three distinct control strategies, and $\xi_{\perp,Z}^2=\frac{N(\Delta\hat{J}_z)^2}{\left|\langle\hat{J}_s\rangle\right|^2}$ is the Wineland squeezing parameter~\cite{3PRA46R6797,4PRA5067}, evaluated for the specific case where the direction of minimal variance, orthogonal to the mean spin, is aligned with the z-axis. Remarkably, our statistics-based GA approach demonstrates a control performance that is not inferior to the highly optimized RL agent~\cite{support}. As shown, after an equivalent number of 600 training iterations, the final squeezing achieved by GA and RL is highly comparable, with both methods significantly outperforming the naive constant control scheme. The final state, visualized by its Husimi Q-function in FIG.~\ref{RLGA}(b1), clearly shows a significant squeezing effect, attesting to the success of the discovered control sequence. 
This outcome underscores the robustness and practical utility of our adaptive GA, positioning it as a potent alternative for quantum control tasks.}

\section{Conclusion}
\label{CONC}
\textcolor{blue}{We have developed and demonstrated a GA framework for preparing spin-squeezed states in open quantum systems. The algorithm's core strength lies in its adaptive crossover and mutation rates, which dynamically balance exploration and exploitation to efficiently navigate the complex control landscape.
We have showcased the versatility of our framework through two distinct optimization tasks: preparing a specific target state with near-unity fidelity ($|\eta|^2 > 0.99$) and directly optimizing for maximum metrological squeezing. A key achievement is not only reaching but also maintaining this high-performance squeezed state over a prolonged time window, a critical feature for practical applications. Furthermore, our analysis reveals the framework's robust scalability. The GA consistently discovers control protocols that yield scalable squeezing, with power-law exponents approaching the fundamental Heisenberg limit, even in the presence of significant dissipation and thermal noise.}

\textcolor{blue}{Critically, through direct comparative analysis, we have demonstrated that our GA-based approach achieves performance competitive with a RL and vastly superior to static control strategies. This result positions the adaptive GA as a potent, robust, and practical alternative in the quantum control toolbox, potentially offering a less parameter-sensitive solution for complex optimization tasks. Given that the proposed control sequences are readily implementable in platforms like atomic Bose-Einstein condensates, and the modular nature of the GA allows for integration with other optimization methods, this work provides a versatile and powerful paradigm for robust quantum state engineering in near-term noisy quantum devices.}

\section{Acknowledgements}
X. L. Zhao thanks the Natural Science Foundation of Shandong Province, China,
No.ZR2020QA078, No.ZR2023MD064, and National Natural Science Foundation of China,
No.12005110, No.12074206. Joint Fund of Natural Science Foundation of Shandong
Province,  No.ZR2022LLZ012. Key Research and Development Program of Shandong Province,
China, No.2023CXGC010901.


\begin{thebibliography}{0}
\bibitem{support}
\textcolor{red}{The code supporting the findings of this study has been deposited in Zenodo under the accession code [DOI: 10.5281/zenodo.17523565].}
\end{thebibliography}
\end{document}